# Abnormal Magnetic Behaviors in Unique Square α-MnO$_2$ Nanotubes

R. Zeng[†,‡], J.Q. Wang[†,∥], W.X. Li[†], G.D. Du[†,‡], Z.X. Chen[†], S. Li[§], Z.P. Guo[†], S.X. Dou

[†]*Institute for Superconducting and Electronic Materials, School of Mechanical, Materials & Mechatronics Engineering, University of Wollongong, NSW 2522, Australia.*

[‡]*Solar Energy Technologies, School of Computing, Engineering and Mathematics, University of Western Sydney, Penrith Sout, Sydney, NSW 2751, Australia*

[∥]*School of Materials Science and Engineering, University of Jinan, Jinan 250022, P. R. China.*

[§]*School of Materials Science and Engineering, University of New South Wales, Sydney NSW 2502, Australia.*

Address for Correspondence:

R. Zeng

Solar Energy Technologies
School of Computing, Engineering and Mathematics
University of Western Sydney
Penrith Sout, Sydney, NSW 2751, Australia
Electronic mail: r.zeng@uws.edu.au

**Abstract**

Systematic magnetic measurements have been performed in unique α–MnO$_2$ square nanotubes synthesized by a facile hydrothermal method with microwave-assisted procedures. Unusual magnetic phenomena (abnormal magnetization verse temperature (M – T) behaviour, large and abnormal magnetization hysteresis loop horizontal shift (H$_{HS}$) verse cooling field (H$_{FC}$) behaviours (H$_{HS}$ – H$_{FC}$) has been observed in these square nanotubes. These suggest the observation of large unfrozen orbital moment which also is the micro-original of observed large and abnormal horizontal shift (H$_{HS}$). The findings demonstrated that engineering layered structures in nanoscale would create many unique nanostructures and unusual physicochemical behaviours.

**Introduction**

Intense experimental and theoretical efforts have been dedicated to understanding the mechanism of nanomagnetism and to form unique magnetic nanostructures. This is due to their huge potential for technological applications in information technology [1, 2] and in other disciplines such as biology and medicine [3]. In last five year, the engineering of layered structures has become more and more important, since the crystallography, electron structure, and physicochemical properties, specially the magnetic properties even the magnetic ground state can change significantly when a layer a few atoms thick is introduced, or even a single atomic layer thickness, such as single carbon atomic layers: graphene [1], and single molecular layers: MnO$_6$ octahedral layers [4], single MoS$_2$ molecular layers [5 - 8], etc. Re-engineering thin layers in different structures could create many more unique structures and present unusual physicochemical phenomena useful for tailoring the properties for applications. Here, we present a study on engineering architectonic MnO$_6$ octahedral molecular layers by restacking thin or single MnO$_6$ layers into alpha-MnO$_2$ tunnel structures linked with MnO$_6$ octahedra and forming a unique square nanotubes, which are interest for their unusual magnetic phenomena. We focus on the formation mechanism of these nanostructures and the relationship between the structures and the unusual magnetic properties (including abnormal magnetization loop horizontal shift (H$_{HS}$) phenomena) by employing systematic microstructure and magnetic measurements and analysis. We demonstrate that, after restacking, the width of the restacking layers (nanoribbons), which account for the nanosheet step-edge, surface, and interface structure of the layers, dominates the magnetic properties of the larger nanostuctures.

MnO$_2$ exist very various crystal phases (α, β, γ, ε, λ, and δ-MnO$_2$). The morphologies of nanostructures in MnO$_2$ are strongly dependent on the preparation conditions (pH value, concentration of cations, and parameters) [9-16, 19, 20]. We have synthesized a series of MnO$_2$ nanostructures with different phases (α, β, ε, and δ) and different nano–architectonic morphologies (nanoflowers, square nanotubes, and tetragonal nanowires) by varying the preparation conditions, the formation mechanism will not discuss here and will appear somewhere else.

In this paper, we present the unusual magnetic phenomena in the unique □□MnO$_2$ square nanotubes. We have performed our magnetism studies with an emphasis on the relationship between microstructures that are mainly on the surface or interface microstructures with magnetic properties, e.g., have a saturated or remanent moment (M$_S$ or M$_R$), and the coercivity (H$_C$), which is significantly enhanced by the step-edges and the surface or interface disordered clusters. We find three main magnetic contributions: a regular antiferromagnetic contribution and two additional irreversible ones under certain conditions. The first contribution can be attributed to the antiferromagnetically ordered tube-wall cores. The nature of the irreversible ones can be identified using DC M–H and M–T curves, AC magnetic curves, thermoremanent (TRM) and isothermoremanent (IRM) magnetization curves, and magnetization hysteresis loop horizontal shift $H_{HS}$–T curves as magnetic identifiers of the irreversible magnetization and spin, and the orbital state. One irreversible feature arises from the spins of conventional uncompensated AFM cores at the interface between the core (AFM α-$MnO_2$) and the spin-glass-like shell. The other one arises from the unusual moment, which is unquenched at low field (< 1 T) and quenched or cancealed at high field. In this case, the moments are antiparallel to the applied field at low field, a phenomenon which seems to disappear at high field, so we argue that there is an unusual high unquenched orbital moment, which is generated by the edge $Mn^{3+}$ orbital moment. In addition, we report the observation of the unusual high unquenched orbital moment and its effects on exchange-bias in α-$MnO_2$ square nanotubes. We show that the TRM/IRM curves combined with the M–T curves, both field-cooled (FC) and zero field-cooled (ZFC), and the exchange-bias versus field and temperature curves can serve generally to identify the irreversible magnetization contribution in a disordered system.

$MnO_2$ nanostructures were synthesized by a facile hydrothermal method with microwave-assisted procedures, which involves the reduction of $KMnO_4$ in a hydrochloric acid solution:

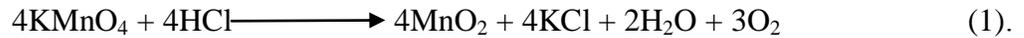

$$4KMnO_4 + 4HCl \longrightarrow 4MnO_2 + 4KCl + 2H_2O + 3O_2 \qquad (1).$$

α–square nanotubes (α–SNT) can be obtained through tuning the microwave irradiation conditions and processing parameters.

The samples have been microstructurally characterized and analysed by x-ray diffraction (XRD) and Rietveld refinement, XPS, scanning and transmission electron microscopy (SEM and TEM), high resolution TEM (HRTEM), Raman spectroscopy, etc. The XRD and Rietveld refinements of α-$MnO_2$ nanotubes (Figure 1(a)) show high phase purity. All samples are single phase nanostructures under the x-ray measurement limitations. However, the XPS result (Fig. 1(b)) shows a small amount of $K^+$ ion including in the sample and moreover, the XPS spectra indicate that the Mn-2p1 peak of sample show splitting to 643.4, 642.7 and 641.5 eV three peak as shown in the insert of Fig. 2(b), which indicates that Mn exists in different valences and oxygen vacancy exists in the surface of the sample. Fig. 2(a) shows field emission SEM (FESEM) images of the $MnO_2$ nanostructures, Fig. 2(b) shows TEM and corresponding selected area electron diffraction pattern (SAED) images, and Fig. 4(1)-(5) shows HRTEM and defect images of $MnO_2$ nanostructures.

The FESEM images show the typical morphologies of α–SNT, those square nanotubes are composed nanosheets of $MnO_6$ octahedral layers associated with layer structure collapse to form tunnel structures with an accompanying phase transformation from δ–phase (layer structure) to α–phase (2 × 2 tunnal). More detailed microstructures and intermediate states of nanostructure formation, and the nanostructure formation mechanism see our another paper eleasewhere. Actually, Fig. 2(b) is a TEM image of single nanotube, a very rough surface and hints at numerous ribbon steps and edges is clear indicated the restacked nanosheets structure. Since α -$MnO_2$ has a tetragonal Hollandite-type structure, in which the $MnO_6$ octahedra are linked to form double zigzag chains along the *c*-axis by edge-sharing, this unique crystal structure may easily form above zigzag step-edges, and, in turn, it can be the origin of the strong unusual ferromagnetism in the nanostructures.

Due to the unique $MnO_6$ octahedral nanoribbons restacked architectonic $MnO_2$ nanotubes, its magnetic behaviors present more unusual features. In order to elucidate those unusual magnetic phenomena, we performed a series of magnetic measurements and analyses, as will be presented, to determine the origin of the magnetism.

Fig. 3 (a) shows M-T curves of α–SNTs after zero field cooling (ZFC) and after field cooling (FC), measured at an applied field of 50 Oe. It shows several unusual characteristic features: (i) there is a ferromagnetic-like transition at $T_C$ = 50 K in this α-MnO$_2$ antiferromagnetic system; (ii) the zero-field cooled $M_{ZFC}$ curve lies above the field cooled $M_{FC}$ one; the difference between the magnetization curves, $\Delta M = M_{FC}-M_{ZFC}$ is also plotted in the figure; and (iii) there is a bifurcation of the FC and ZFC magnetizations at a so-called blocking temperature $T_B = T_C$, which is a sign of irreversible contributions.

We collected ZFC and FC M-T curves (as shown in Fig. S1(a)) with measurements at different fields: 50 Oe, 100 Oe, 500 Oe, 1 kOe, 5 kOe, 1 T, 2 T, 3 T, and 5 T. Their $\Delta M$ (= $M_{FC}-M_{ZFC}$) – T curves are shown in Fig. 3(b), and their derivative curves ($\delta M/\delta T$ – T) curves are shown in Fig. S1(b). From these curves, we can ascertain that there are three obvious magnetic transitions and one hinted transition in this system over the measured temperature range. A kink in both the FC and the ZFC curves is found, which usually signals simple AFM behavior. (The $T_N$ of nanosized α-MnO$_2$ is about 13 K [17, 18].) The derivative curves ($\delta M/\delta T$ – T) more clearly show these transitions, as marked in the figure: the $T_N$ of the nanoribbon based α–MnO$_2$ (at about 13 K) certainly accompanies AFM features, e.g., there is a slight shift to higher temperature with increasing field; spin-glass-like (SG) or superparamagnetic (SPM) behavior of the peak temperature $T_{Peak}$ is observed at about 40 K, which shows a strong shift with increasing field; while the above-mentioned ferromagnetic-like transition at $T_C$ is at about 50 K, that does not shift or only very slightly shifts with the applied field. In the difference between the magnetization curves, $\Delta M = M_{FC}-M_{ZFC}$, as presented in Fig. 3(b), which only shows the irreversible contributions, we find three features: (i) the curves are monotonically decreasing with temperature, reflecting the expected thermally induced decay of the magnetization ; (ii) there are unusual negative $\Delta M$ values (since the zero-field cooled $M_{ZFC}$ curve lies above the field-cooled $M_{FC}$ one); and (iii) there are negative $\Delta M$ values that monotonically decrease with field and approach zero at 1 T, while there is a return to the normal state (positive $\Delta M$ values) at fields higher than 1 T, which reflects an unexpected field induced transition in the reversible magnetization and competition resulting in a critical cooling field $H^R_C \approx$ 1 T.

In order to better explain the features, the AC susceptibility ($\chi'$) vs. T curves and the reversible moment analysis are presented in Fig. S1 (c) and (d), they shows $\chi'$–T curves under different frequency (f) and field ($H_p$), respectively. The $\chi'$ reflects the reversible moment, and it can be seen that the $\chi'$ value slightly changes with f and strongly changes with $H_p$, and it is more importantly indicated that the peak position slightly shift with the f and $H_p$. The HRTEM observations present numerous α-Mn$_2$O$_3$, or β-MnO$_2$, etc. heterostructural clusters (of a few nanometers) dispersed in the surface, especially near the edges (in Fig. 2(d)). We are very sure that these clusters cause a SPM cluster-like behavior, because the peak position ($T_{Peak}$) strongly shifts with increasing field, a feature of SPM behavior [21-24].

The unusual behavior in the $\Delta M$ (= $M_{FC}-M_{ZFC}$) – T curves (Fig. 3(b)), with the $\Delta M$ values displaying changes from negative to positive, strongly hints at a third magnetic subsystem – most probably the edges and steps containing Mn$^{3+}$ ions of α–MnO$_2$ nanoribbons and heterostructural clusters, which have strongly unquenched orbital moment and strongly interactions with each other, so that they have the same critical transition temperature, which may be another type of proximity effect feature [25] in these unique square α-MnO$_2$ nanotubes, since the $T_N$ of those heterostructural clusters (such as β–MnO$_2$) may be higher than for the α-MnO$_2$ nanotubes. Oxygen vacancies and uncompensated electrons of edge Mn ions cause Mn$^{3+}$/ Mn$^{4+}$ exchanges (due to the large orbital moment in Mn$^{3+}$ [26]). XPS measurements indicated that there is splitting of the Mn 2p peak, along with a small amount of Mn$^{3+}$ ions, while a K 1s peak is present. The K$^+$ cations that are present are essential for the formation and stabilization of the α–MnO$_2$ nanostructure, and inevitably, the introduction of $K^+$ cations into the tunnel cavity causes a mixture of $Mn^{3+}$ / $Mn^{4+}$ in the system based on the valence balance. The $Mn^{3+}$ ions at step-edges give rise to well aligned, long range ordered FM or ferrimagnetism (FIM), and they have high anisotropy energy due to the very high orbital moment [27-29]. These couple with the $Mn^{3+}$/ $Mn^{4+}$ (as the spin moment is antiferromagnetic) in the bulk and are the origin of spin and orbital coupling, as well as competition, so the orbital moment is ultimately responsible for the unusual magnetic behavior.

In order to better understand this unusual magnetic behavior, further magnetic studies were performed. M - H hysteresis loops measured at temperatures of 5, 10, 30, 50, and 70 K after ZFC with applied fields up to 90 kOe are shown in Fig. S2, the inset shows the $H_C$ vs. temperature curve. To gain further insight into the origin of this small net magnetization of α-$MnO_2$ nanotubes, hysteresis loops cooled under different fields have been collected (Fig. S3(a). The M-H hysteresis loop may display an enhancement of the coercive field ($H_C$), and an vertical or horizontal shift due to the interfacial coupling effects when field cooling the samples lower than blocking temperature ($T_B$), e.g. the FC M-H hysteresis loops may exhibit the typical exchange bias - like features, namely a horizontal shift of the hysteresis loop ($H_{HS}$) due to the interface of core-shell magnetic coupling. Although the displayed shifts may include two contributions: (i) unsaturation remanence magnetisation when the applied filed lower than the so-called reversal field [30, 31] (ii) the real exchange bias ($H_{EB}$). In order to consistent with the exchange bias research community, we here donate the horizontal shift as $H_{HS}$, since we didn't determine the reversal field yet. We have systematically measured magnetization hysteresis loops after ZFC and FC on α-$MnO_2$ square nanotubes, the summary results, are shown in Fig. 4(a). $H_{C1}$, $H_{C2}$, and $H_C$ at T = 5 K under different cooling fields have been measured for the α-$MnO_2$ nanotube sample. Four curves of $H_{C1}$, $H_{C2}$, $H_{HS}$, and $H_C$ versus cooling field ($H_{FC}$) are shown in Fig. 4(a). A strong dependence of the $H_{FC}$ on the exchange bias magnitude is observed, and we note that there are different tendencies for $H_{C1}$ and $H_{C2}$: $H_{C1}$ monotonically increases with $H_{FC}$, but $H_{C2}$ first increases with $H_{FC}$ along with $H_{C1}$ before about 1 T (again $H^R_C \approx 1$ T) and then changes to decrease in the opposite direction, which is another piece of evidence for the existence of a superferromagnetic (SFM) phase and step-edges in the square nanotubes. Surprisingly, we have observed that the $H_{HS}$ rapidly increases under cooling fields of less than about 1 T and then slowly decreases with further increases in the cooling field. Changes in the $H_C$ show an opposite tendency with $H_{HS}$. Moreover, selected M – H hysteresis loops at different temperatures after 2 T $H_{FC}$ cooling have been collected (as shown in supplementary Fig. S4(b)), and four curves of $H_{C1}$, $H_{C2}$, $H_{HS}$, and $H_C$ versus temperature after 2 T $H_{FC}$ are shown in Fig. 4(b). Strong temperature dependence of the $H_{HS}$ magnitude is observed as well, and we note that there is different behavior for $H_{C1}$ and $H_{C2}$: $H_{C1}$ monotonically decreases with increasing T, but $H_{C2}$ first increases with increasing T before ~ 13 K (the $T_N$ of nano-α-$MnO_2$), and then decreases, while $H_{C1}$ decreases in a linear way with increasing T. The detailed analysis see the supplementary information. In our system, the $H_C$ peak and the positive $H_{HS}$ appear at different temperature points, i.e., the $H_C$ peak appears at around $T_N$ and the positive $H_{HS}$ at around $T_C$, which further indicate and confirm that there are multiple exchange couplings in our system, e.g cyclic exchange coupling and competition among AFM/SG, SG/FM, AFM/SG and AFM/FM, but there is a dominant type of coupling for the different temperature ranges.

In order to further test this idea, we performed measurements of the TRM / IRM vs. H at 5 K in the field range of 50 Oe < H < 90 kOe, as shown in Fig. 5(a), and TRM / IRM vs. T under 2 T field, as shown in Fig. 5(b). The TRM was measured under the following conditions: the system was cooled in the specified field from room temperature down to 5 K, the field was removed, and then the magnetization was measured. The IRM was measured under the following conditions: the sample was cooled in zero field from room temperature down to 5 K, the field was then momentarily applied, removed again, and the remnant magnetization measured. It is important to note that the TRM and the IRM probe two different states. The TRM probes the remnant magnetization in zero field after freezing in a certain magnetization in an applied field during FC. However, the

IRM probes the remnant magnetization in zero field after ZFC (in a demagnetized state) and after magnetizing the system at low temperatures. Therefore, systems with a nontrivial H-T-phase diagram will show a characteristic difference between TRM and IRM. First, we analyze the TRM / IRM - H curves (Fig. 5(a)), M.J. Benitez et al. [41] have listed the characteristic shapes for three different systems of TRM / IRM - H curves and suggested using these curves as magnetic fingerprints of irreversible magnetization and as identifiers of the type of system. For example, the spin-glass state strongly depends on whether it is cooled in a field or not, and Fig. 5(a) shows similar behavior to that of the spin-glass state, which further confirms our arguments above. Fig. 5(a) shows the TRM and IRM curves as function of magnetic field for our $MnO_2$ nanotubes, which resembles the canonical spin-glass system [27-29, 32]. However, to further make the case for a spin glass, two features can be observed in Fig 5(a). First, the IRM increases relatively strongly with increasing field and then meets the TRM curve at a moderate field value, where both of them then saturate. Second, the TRM exhibits a characteristic peak at intermediate field, which is also reproduced in several other studies found in the literature [33]. A superparamagnetic system shows a qualitatively similar plot, however, without this characteristic peak in the TRM curve [34]. We find similarities in the behavior of our $MnO_2$ nanotubes to both spin-glass behavior and the behavior of a diluted AFM in a field (DAFF) system. However, in our case, the TRM does not appear to obviously show two peaks or kinks; hence, one can exclude the possibility of superparamagnetic behavior; for the IRM, however, there is different behavior, e.g. the IRM has negative values when the applied field is less than 1 T, which agrees with the conclusion from the M-T curves that the critical cooling field field $H^R_C \approx 1$ T exists, as mentioned above. On comparing the TRM/IRM plot to the spin-glass and the DAFF systems, we find good correspondence to the DAFF system when the $H_{FC}$ is less than 1 T and very good correspondence to the spin-glass system when the $H_{FC}$ is larger than 1 T. We argue that the TRM/ IRM behavior of the α-$MnO_2$ square nanotubes corresponds to both the DAFF and the spin glass.

The TRM / IRM – T curves presented in Fig. 5(b) seem to show two blocking-like temperatures, one is $T_{BSG}$, which is close to $T_{SG}$, and one is $T_{BC}$, which close to $T_C$ under a 2 T cooling field. These further confirm our arguments for multiple exchange couplings in our system. Comparing the $H_C$ – T curves (Fig. 4(b)), we note that the $H_C$ peak appears at $T = T_{BSG} = T_{Peak}$, but the $H_{HS}$ – T curve changes smoothly at this temperature. In addition, the different behavior of $H_{HS}$ and $H_C$ with $H_{FC}$ indicates that the mechanism responsible for $H_C$ enhancement is independent of the origin of $H_{HS}$. From these findings, we can conclude that the magnitude of the pinned uncompensated interfacial AFM moments that give rise to the exchange bias depend on not only the interfacial spins, but also the entire bulk AFM magnetic structure. These allow us to control the $H_{EB}$ magnitude by changing not only the cooling conditions, also the system morphology.

From these findings, we conclude that the nanoribbons stacked into α – SNTs consist of AFM ordered ribbon cores, which behave as AFM materials, while the shell surfaces or interfaces are different from the α–$MnO_2$ structural clusters and step-edges, so they interact with each other, which produces SPM, DAFF, spin-glass-like, and FM behaviors. In a SPM system, the peak positions show a much stronger shift with increasing field [21 - 24], and the M – T curves under field from 50 Oe to 5 T (Fig. 4(a) and (b)) show a strong shift with increasing field. Following the detailed magnetic analysis on α – SNTs, we have systemically analysed the other $MnO_2$ nanostructures presented in this paper. Due to their different surface or interface microstructures, they show slightly different magnetic behaviours, so that only AFM/SG interface magnetic exchange coupling behaviour is observed in core-shell structural nanoribbons of δ–NF, and these results will be presented elsewhere. However, the microscopic origins of magnetization (M) and coercivity ($H_C$) are similar. It is well known that disorder from element vacancies, valence changes, defects and strains, zigzag edges, and even thermal effects [35-41], etc. can result in the formation of random clusters that induce weak magnetism in nanostructures, but the same can arise from well aligned structures, such as step-edges, which can create strong magnetic anisotropy [26 - 29]. All of these may exist in our $MnO_2$ nanostructures, especially in the surfaces and interfaces of the nanoribbons. Restacking of those nanoribbons leads to coexistence and competition between different magnetic behaviours and would strongly enhance their interaction exchange couplings, which generate the unusual magnetic phenomena.

The magnetic $Mn^{3+}$/$Mn^{4+}$ ions are triangularly arranged, which indicates that strong geometrical frustration (GF) may exist in the sample. Such frustration is one of the main causes for the occurrence of a spin-glass-like (SG-like) state, which is confirmed by our above magnetic measurements and analysis (in Fig. 3(b) Fig. S2(c), Fig. 5 (a)), and Fig. S5(a, b)). The SG-like behavior may be due to the randomly arranged (or disordered) GFs due to $K^+$ ions or surface/interface random $Mn^{3+}$ clusters caused by oxygen vacancies (see Fig. 1). If these GFs give rise to well-ordered structures, such as (i) a well-aligned line, which may cause chiral magnetic order [42], (ii) a skyrmion lattice-like ground state [43], and even (iii) a lattice structure similar to the $H^+$ positions in ice (frozen water), which may artificially create magnetic monopoles of spin ice [44], etc., unusual magnetic phenomena would be observed. The atomic arrangements in the crystal and thus the magnetic GFs that are possibly different at the step-edges and corners have been analyzed and modeled. Fig. 6 shows the crystal structures of the α-$MnO_2$ sample with only Mn ions indicated. It displays clearly the triangular lattice configuration of magnetic atoms at a (1) step-edge; (2) step-edge with outside corner; and (3) step-edge with inside corner. The possible magnetic moment array of the above step-edge (i) and its outer (ii) and inner (iii) corners may be aligned as in the arrangements shown in Fig. 6(d) with the additional scheme (iv) of four inner corners and outer corners in a cross-section of a square nanotube. This means that the GFs may be well aligned and well ordered to different types of states in the square nanotubes, which could create the above-mentioned chiral skyrmion lattice-like and even spin ice ground states, since the $Mn^{3+}$ has a high orbital moment [45] and it create a GF with $Mn^{4+}$, which indicates that GFs may also exist with high spin-orbital coupling, in addition to the situation at step-edges, etc., which all create favorable conditions for chiral skyrmion lattice formation [43]. The square magnetic GF array in the cross-section of a nanotube would form a spin-ice-like lattice [44].

In the real system, the above-mentioned (i) to (iv) magnetic lattice arrays may exist over short distances, but there are long-distance interaction, with the competition being further excited and time-relaxed, depending on the intrinsic system balance, and extra thermal and field excitation. These lead to the complication of macroscopic magnetic phenomena that are observed by conventional magnetic measurements, but which can be used to interpret the observation of unusual magnetic phenomena very well. For examples, the negative ΔM (= $M_{FC} - M_{ZFC}$) value and its feature of $H^R_C$ (in Fig. 6 and Fig.S4) may due to the chiral skyrmion lattice formation, which generates vortex-like moments antiparallel to the field direction. The critical field to reverse this moment is $H^R_C$, but the phenomenon also can be simply interpreted by the formation of a "spin-ice"–like lattice, with frozen monopole moments due to field cooling, which could cause enormous uniaxial magnetocrystallographic anisotropy. This also provides a simple explanation for the abnormal exchange bias phenomena. It is the processing of the field that eliminates those moments when the field is lower than $H^R_C$; under high field, these moments disappear or are reversed, so the exchange interaction ($H_{EB}$) would be eliminated as well, but the $H_C$ retains the same value (see Fig. 4).

In summary, we present studies on a facile microwave-assisted hydrothermal synthesis method to prepare nanoribbons a few atoms thick, which stack into unique $MnO_2$ square nanotubes. The unique morphologies hence present unusual magnetic phenomena. In the surface or in the interface, the $MnO_2$ heteroclusters and step-edges, which are associated with variation of the valence of Mn ions, are the microscopic origin of the ferromagnetism, and their interactions, couplings, and competition cause the unusual magnetic phenomena.

We find multiple magnetic contributions in the nanostructures: (i) regular antiferromagnetically ordered nanostructure cores; (ii) surface/interface heterostructural clusters or shells with $Mn^{3+}$/$Mn^{4+}$ coupling, which contribute to the magnetic moments and present spin-glass-like magnetic behaviors; (iii) long-range oriented nanoribbons presenting restacking step-edges with oxygen vacancies (more $Mn^{3+}$ ions), defects, and strains attributable to highly anisotropic ferromagnetic moments, and the features of high coercivity fields and irreversible magnetic moments. Special investigations have been performed on the surprising and unusual magnetic phenomena in α-$MnO_2$ square nanotubes, where it was found that some magnetic moments (spins) antiparallel to the applied field direction exist in square nanotubes, we argue that this may suggests the observation of an unquenched moment or chiral skyrmion-like moments or even spin-ice moments, further intensity study required to confirm this point. The microscopic origin of this moment may due to $Mn^{3+}$ ion orbital moments and strong spin-orbital coupling, which causes chiral skyrmion-like moments at well aligned step-edges, especially in the corners of step-edges.

Although these claims and interpretations are drawn only for these particular square nanotubes, they might still be applicable to many emerging materials, including artificially or self-assembled nanostructure arrays, especially for artificial nanostructures with layer stacking, such as graphene layer stacking, and for the interpretation of results on other complicated experimental magnetic nanosystems.

The authors thank Dr. T. Silver for her help and useful discussions. This work is supported by the Australian Research Council.

**Figure Captions**

Fig. 1. (a) XRD patterns with Rietveld refinement results on α-MnO$_2$ square nanotubes (α-SNT). The crystal structure of α-SNT is tetragonal with space group I4/m, and the the lattice parameters are $a = b = 9.85$ Å, $c = 2.86$ Å, and (b) XPS spectra of α-SNT.

Fig. 2. (a) FESEM image (b) TEM image, (c) the schemes image of single nanosheets stacked square nanotube and (d) HRTEM of a single nanotube of α–MnO$_2$ square nanotubes. Details of the insets and composite panels are given in the text.

Fig. 3. M vs. T curves of α-MnO$_2$ square nanotubes after zero field cooling (ZFC) and after field cooling (FC), measured at different DC and AC applied fields, i.e., M-T under a field of 50 Oe (a) and ΔM (= $M_{FC}$-$M_{ZFC}$)-T curves under DC fields of 0.005, 0.05, 0.1, 0.5, 1, 2, 3, and 5 T (b).

Fig. 4. Relationships of abnormal magnetization hysteresis loop horizontal shift ($H_{HS}$) and coercivity field ($H_C$) of α-MnO$_2$ square nanotubes under different values of field cooling ($H_{FC}$) and at different temperatures: (a) plots of the recorded abnormal magnetization hysteresis loop horizontal shift ($H_{HS} = (H_{C1} - H_{C2})/2$ ) and coercivity field ($H_C = (H_{C1} + H_{C2})/2$) under different $H_{FC}$; $H_{C1}$, and $H_{C2}$ vs. $H_{FC}$ are also plotted alongside; (b) plots of the recorded magnetization hysteresis loop horizontal shift ($H_{HS}$) and coercivity field ($H_C$) under different $H_{FC}$; $H_{C1}$ and $H_{C2}$ versus temperature are also plotted alongside, while the inset plots an enlargement of the figure near the $T_C$.

Fig. 5. TRM and IRM features of α-MnO$_2$ square nanotubes under different $H_{FC}$ and different temperatures: (1) TRM and IRM vs. $H_{FC}$ at 5 K; (2) TRM/IRM vs. T plots after $H_{FC} = 2$ T field cooling.

# Figures

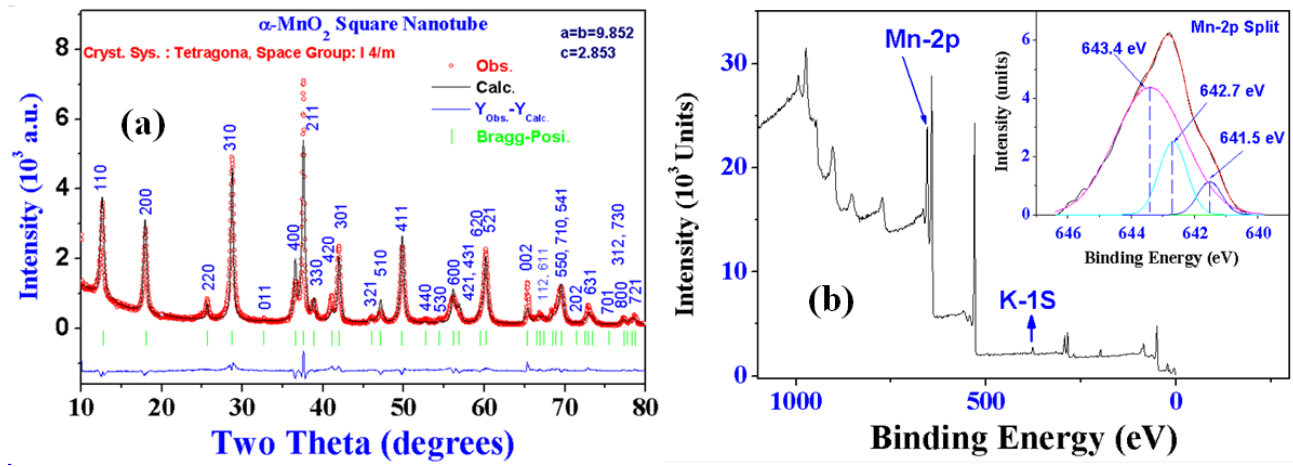

Fig. 1, R. Zeng et al.

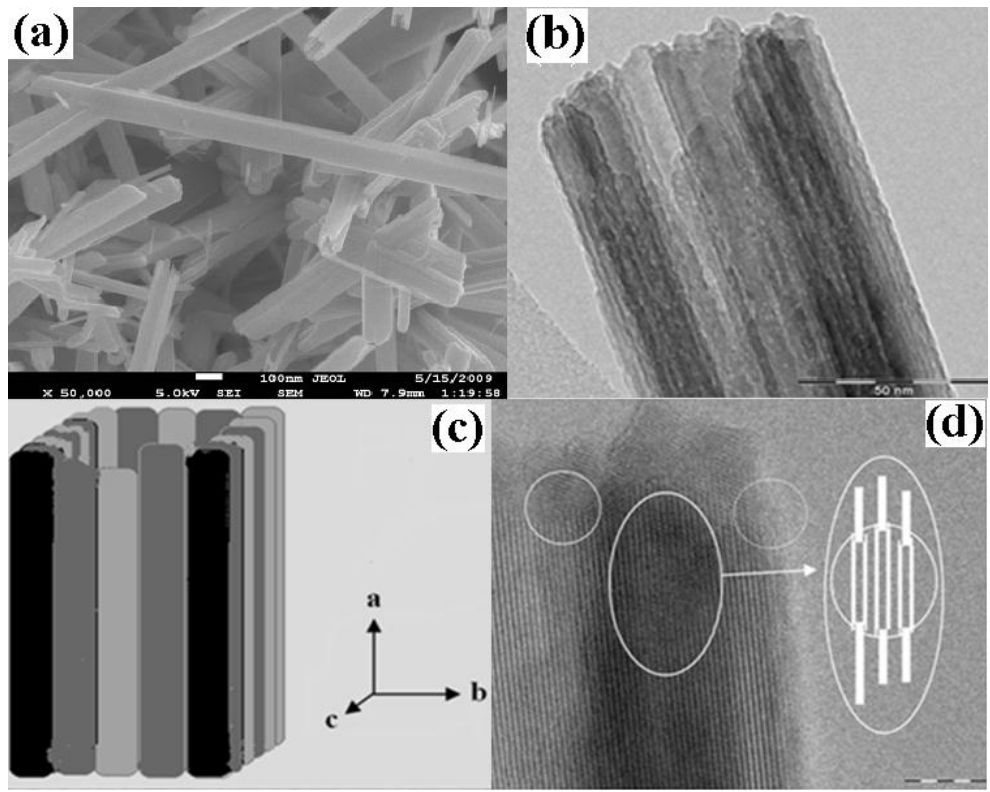

Fig. 2, R. Zeng et al.

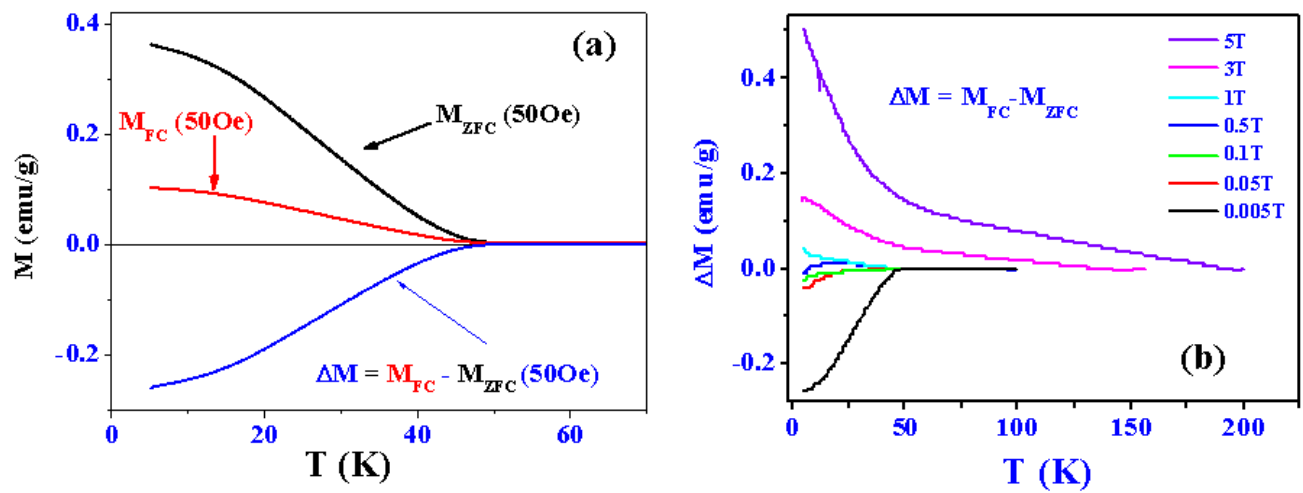

Fig. 3, R. Zeng et al.

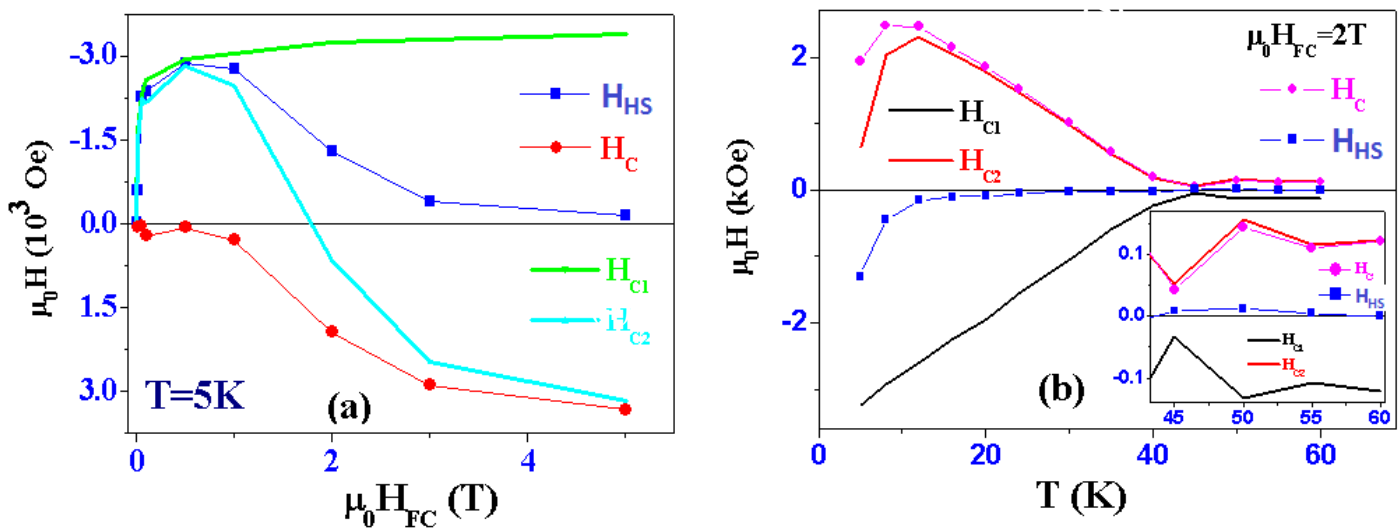

Fig. 4, R. Zeng et al.

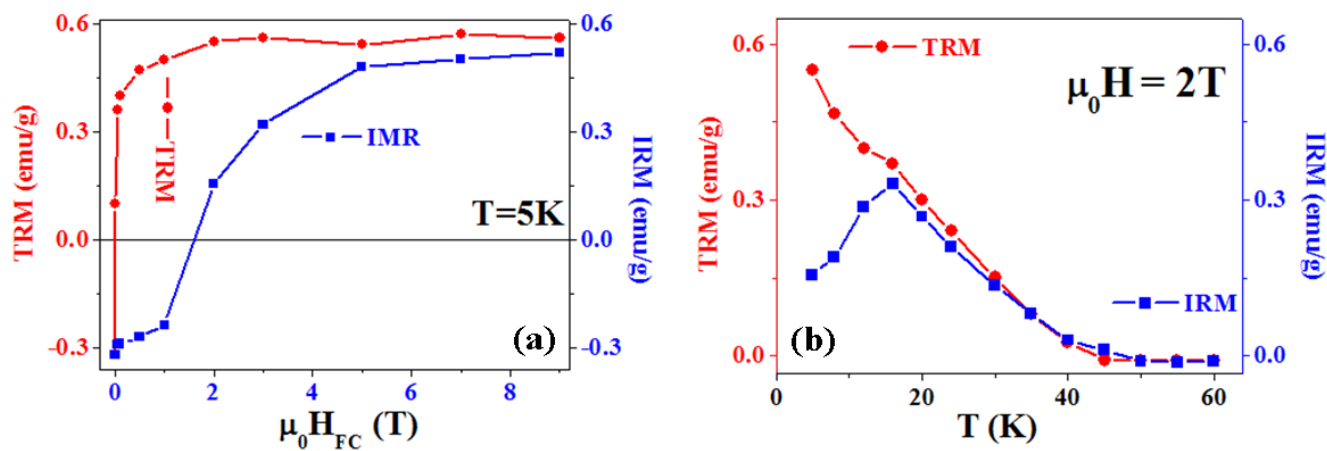

Fig. 5, R. Zeng et al.

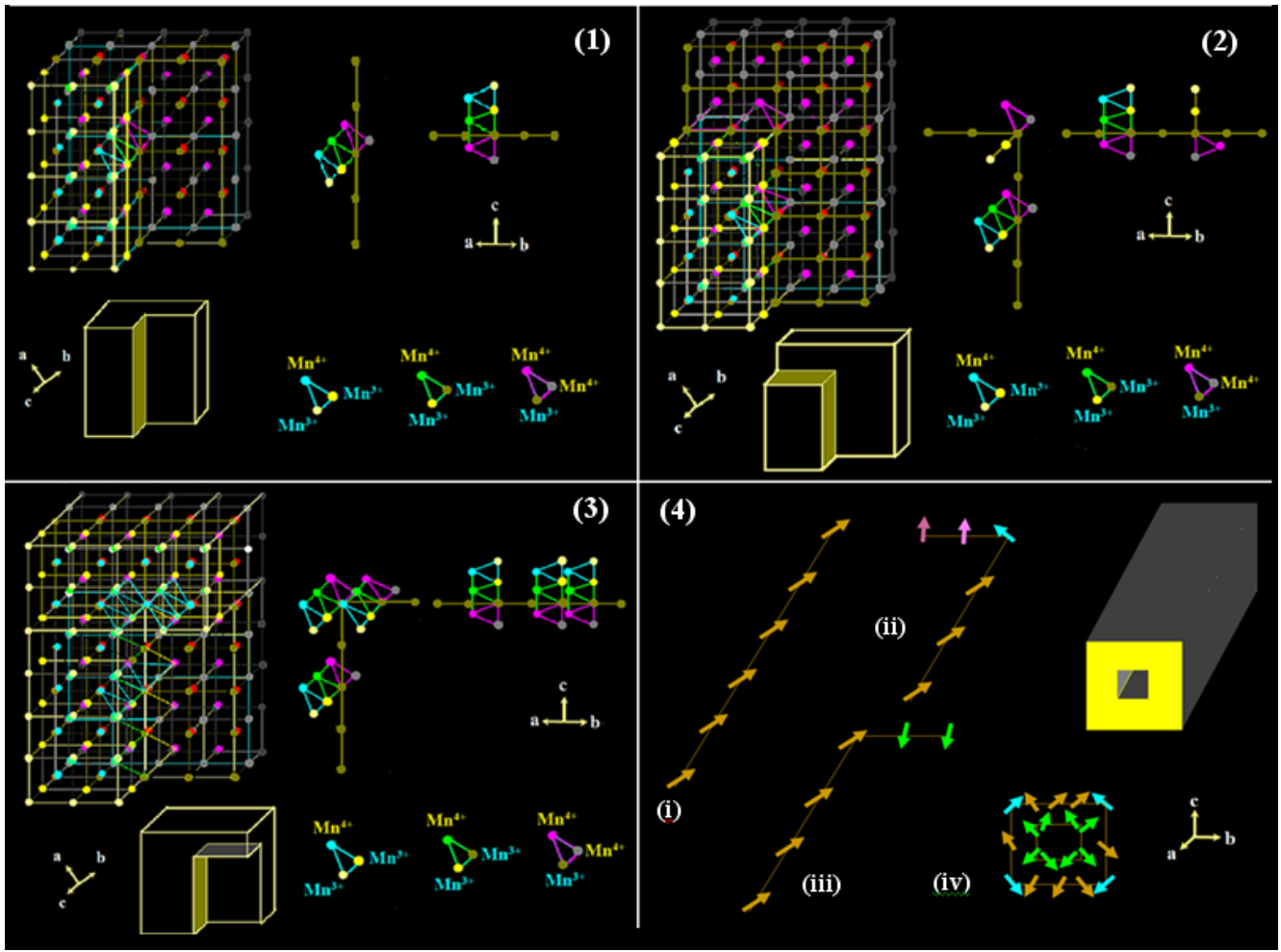

Fig. 6, R. Zeng et al.

# Supporting Information

# for

# Abnormal Magnetic Behaviors in Unique Square α-MnO$_2$ Nanotubes

R. Zeng[†,‡], J.Q. Wang[†,∥], W.X. Li[†], G.D. Du[†,‡], Z.X. Chen[†], S. Li[§], Z.P. Guo[†], S.X. Dou

[†]*Institute for Superconducting and Electronic Materials, School of Mechanical, Materials & Mechatronics Engineering, University of Wollongong, NSW 2522, Australia.*

[‡]*Solar Energy Technologies, School of Computing, Engineering and Mathematics, University of Western Sydney, Penrith Sout, Sydney, NSW 2751, Australia*

[∥]*School of Materials Science and Engineering, University of Jinan, Jinan 250022, P. R. China.*

[§]*School of Materials Science and Engineering, University of New South Wales, Sydney NSW 2502, Australia.*

Address for Correspondence:

R. Zeng

Solar Energy Technologies
School of Computing, Engineering and Mathematics
University of Western Sydney
Penrith Sout, Sydney, NSW 2751, Australia
Electronic mail: r.zeng@uws.edu.au

Over the past five years, single layer carbon graphene nanoribbons have attracted the greatest interest among nanomaterial researchers. However, some of the properties of carbon-based materials are not suitable for all applications. For example, graphene nanoribbons exhibit different electronic and magnetic properties, depending on the chirality of their edges [1, 2]. Magnetic nanostructures and nanomagnetism, however, have always attracted much interest among magnetism researchers. This is due to their huge potential for technological applications in information technology [3, 4] and in other disciplines such as biology and medicine [5]. A challenging aim of current research in magnetism is to explore structures of still lower dimensionality, [6-8] and to explore the spins, orbital lattices, and couplings in the low dimension nanostructures [9-11]. As the dimensionality of a physical system is reduced, magnetic ordering tends to decrease, as fluctuations become relatively more important, but it seems that this can be overcome by engineering the surface nanostructures and step-edge atoms through introducing exchange bias [11] and enhancing the magnetic anisotropy, since step atoms present a remarkably high anisotropy energy in two-dimensional nanostructures [12-13].

In particular, nanostructures consisting of an antiferromagnetic (AFM) material have been of most interest in the last five years [14]. As the size of a magnetic system decreases, the significance of the surface and its roughness or surface step atom quantum effects increases. Since an antiferromagnet usually has two mutually compensating sublattices, the surface always leads to a breaking of the sublattice pairing and thus to ''uncompensated'' surface spins. This effect has already been explained as the origin of exchange-bias or the enhancement of magnetic anisotropy, and net magnetic moment in AFM nanoparticles by Neel [15].

Several experimental studies followed, suggesting various scenarios for the magnetic properties found, e.g., surface roughness and surface step atoms make remarkably different contributions to magnetic moment, spin-glass or cluster-glass-like behavior of the surface spins [16-19], thermal excitation of spin-precession modes [20], finite-size induced multi-sublattice ordering [21], core-shell interactions [22-26], or weak ferromagnetism [27, 28]. However, precise identification of the nature of the surface contribution has remained unclear. Terms such as ''disordered surface state,'' ''loose surface spins,'' ''uncoupled spins,'' ''spin-glass-like behavior,'' "high anisotropy energy step-edge orbital moment," etc. express the uncertainty in the description of the surface contribution.

In order to investigate and attain a deeper understanding of the surface spin or the rough surface step contributions in AFM nanosystems, we have synthesized high-quality AFM $MnO_2$ nanostructures. Manganese dioxides with tunnel structures are attractive inorganic materials owing to their distinctive physical and chemical properties, as well as their wide applications in molecular/ion sieves, [3, 29] catalysts, [30, 31] and electrode materials in $Li/MnO_2$ batteries. [32-35]. Even their unclear response mechanism has hugely improved the sensitivity of biosensors [36]. Over the past few years, various $MnO_2$ nanostructures with different morphologies and crystallographic forms have been reported. [37-39] $MnO_2$ has many polymorphic forms, such as α, β, γ, ε, λ, and δ-$MnO_2$, which are different in the way they link the basic $MnO_6$ octahedral units [40, 41]. Among them, α-$MnO_2$ for example, has a tetragonal Hollandite-type structure with the space group *I*4/*m*, in which the $MnO_6$ octahedra are linked to form double zigzag chains along the *c*-axis by edge-sharing. These double chains then share their corners with each other to form approximately square tunnels parallel to the *c*-axis. Since the tunnel cavity is as large as 0.46 nm, it is therefore inevitable that some large cations such as $K^+$, $Ba^{2+}$, and others are introduced into the tunnel from the raw materials during the synthesis process, which could adjust the Mn valence and bond distances or cause lattice strain, hence influencing the magnetic and other properties. In addition, the tetragonal rutile-type β-$MnO_2$ is the thermodynamically most stable and abundant member in the manganese dioxide family, and it plays an important role in magnetism and transport properties [40,41]. It is well known that β-$MnO_2$ shows a magnetic transition into a helical state at the Neel temperature ($T_N$) of about 92 K, below which it has a well-known screw type incommensurate magnetic structure, with the pitch of the screw about 4% shorter than 7/2$c$ [42]. Above $T_N$, the magnetoresistance (MR) of β-MnO$_2$ is slightly negative and isotropic. However, below $T_N$, on the other hand, the MR becomes anisotropic and remains small. In this report, we focus on the magnetic behaviors and transport properties of magnesium dioxide. The electrochemical properties will be presented in other papers.

It is well known that α-MnO$_2$ is antiferromagnetic, but it is possible that the insertion of large cations into the tunnel in the α-MnO$_2$ structure could lead to changes in its magnetic properties. On the one hand, a reduction in the oxidation state of the Mn mixed valence of Mn$^{3+}$ and Mn$^{4+}$ is necessary in order to compensate the charge of the introduced large cations, [29] which could influence the magnetic coupling between Mn cations. On the other hand, the distribution of the Mn$^{3+}$ and Mn$^{4+}$ cations should be closely related to the distribution of the intercalated cations, which in turn may cause a change in the magnetic ground state. For example, spin-glass (SG) behaviour has been observed in KMn$_8$O$_{16}$ compound with the same crystal structure as α-MnO$_2$.

Here we focus on the magnetic measurements on α-MnO$_2$ square nanotubes. Basic magnetic properties of the sample have been measured using a commercial vibrating sample magnetometer (VSM) model magnetic properties measurement system (MPMS) magnetometer (Quantum Design, 14 T), in applied magnetic fields up to 70 kOe. The ZFC and FC M-T curves with measurements at different fields: 50 Oe, 100 Oe, 500 Oe, 1 kOe, 5 kOe, 1 T, 2 T, 3 T, and 5 T are shown in Fig. S1 (a). The derivative curves ($\delta M/\delta T - T$) curves are shown in Fig. S1(b). The peak position in the ZFC curve marks the onset of AFM long-range order and is also often considered to mark the critical temperature $T_N$. Two other transition temperatures are marked in Fig. S1(b), $T_C$ and $T_{Peak}$. The hinted transition is a field induced ferromagnetic (FM) to paramagnetic (PM) transition or field suppressed ferromagnetism transition, since neither the M – T curves (Fig. S1(a)) nor the $\delta M/\delta T - T$ curves (Fig. S1(b)) show the $T_C$ transition when the field value is higher than 1 T.

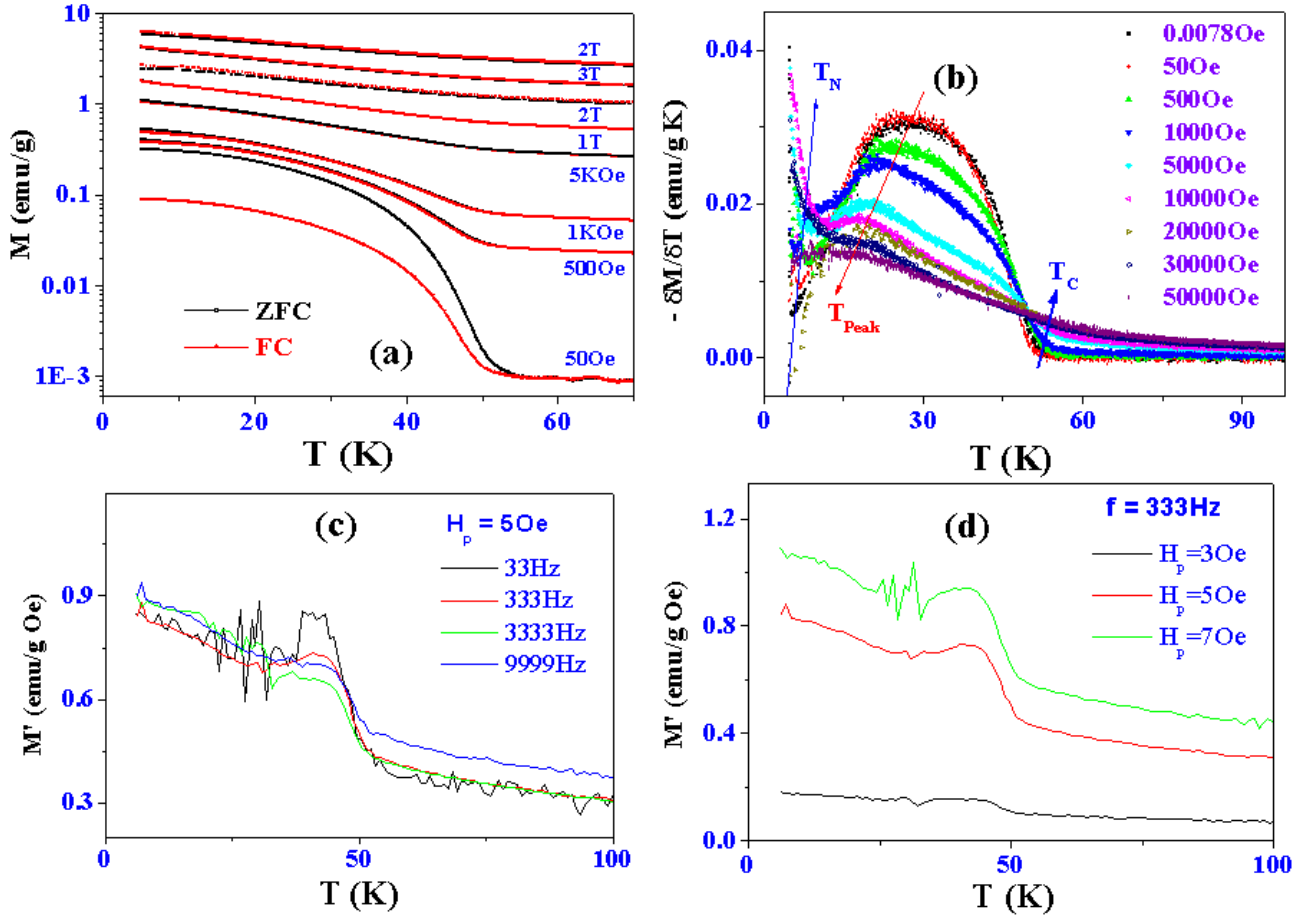

**Fig. S1.** M vs. T curves of α-MnO$_2$ square nanotubes after zero field cooling (ZFC) and after field cooling (FC), measured at different DC and AC applied fields, i.e., under different fields from 50 Oe to 50 kOe **(a)**; derivative δM/δT – T (ZFC) curves under DC fields of 0, 0.05, 0.1, 0.5, 1, 2, 3, 5T: $T_C$, $T_N$, and $T_{Peak}$ are marked by arrows **(b)**; AC susceptibility M′ – T curves after ZFC under $H_P$ = 5 Oe and different frequencies f = 33, 333, 3333, and 9999 Hz **(c)** and under f = 333 Hz at different $H_P$ = 3, 5, 7 Oe **(d)**.

The derivative curves (δM/δT – T) more clearly show these transitions, as marked in the figure: the $T_N$ of the nanoribbon based α–MnO$_2$ (at about 13 K) certainly accompanies AFM features, e.g., there is a slight shift to higher temperature with increasing field; spin-glass-like (SG) and superparamagnetic (SPM) behavior of the peak temperature $T_{Peak}$ is observed at about 40 K, which shows a strong shift with increasing field; while the above-mentioned ferromagnetic-like transition at $T_C$ is at about 50 K, which does not shift or only very slightly shifts with the applied field.

Fig. S1(c) and S1(d) shows χ′–T curves under different frequency (f) and field ($H_p$), respectively. The χ′ reflects the reversible moment, and it can be seen that the χ′ value slightly changes with f and strongly changes with $H_p$, and it is more importantly indicated that the peak position slightly shift with the f and $H_p$. Also, the thermoremanent magnetization (TRM) does not show a training effect with temperature, and the isothermoremanent magnetization (IRM) does not show a training effect with field and time, while the magnetization hysteresis loop horizontal shift ($H_{HS}$) and TRM have a training effect with cycle number (supporting information Fig. S4(a)), and the TRM/IRM vs. cooling field ($H_{FC}$) and temperature behavior, which will be discussed later, is a sure indication of spin-glass (SG) cluster-like behavior in the system. On the other hand, our HRTEM observations present numerous α-Mn$_2$O$_3$, β-MnO$_2$, etc. heterostructural clusters (of a few nanometers) dispersed in the surface, especially near the edges (see Fig. 2(d). We are very sure that these clusters cause a SPM cluster-like behavior, because the peak position (T$_{Peak}$) strongly shifts with increasing field, a feature of SPM behavior [43 - 46].

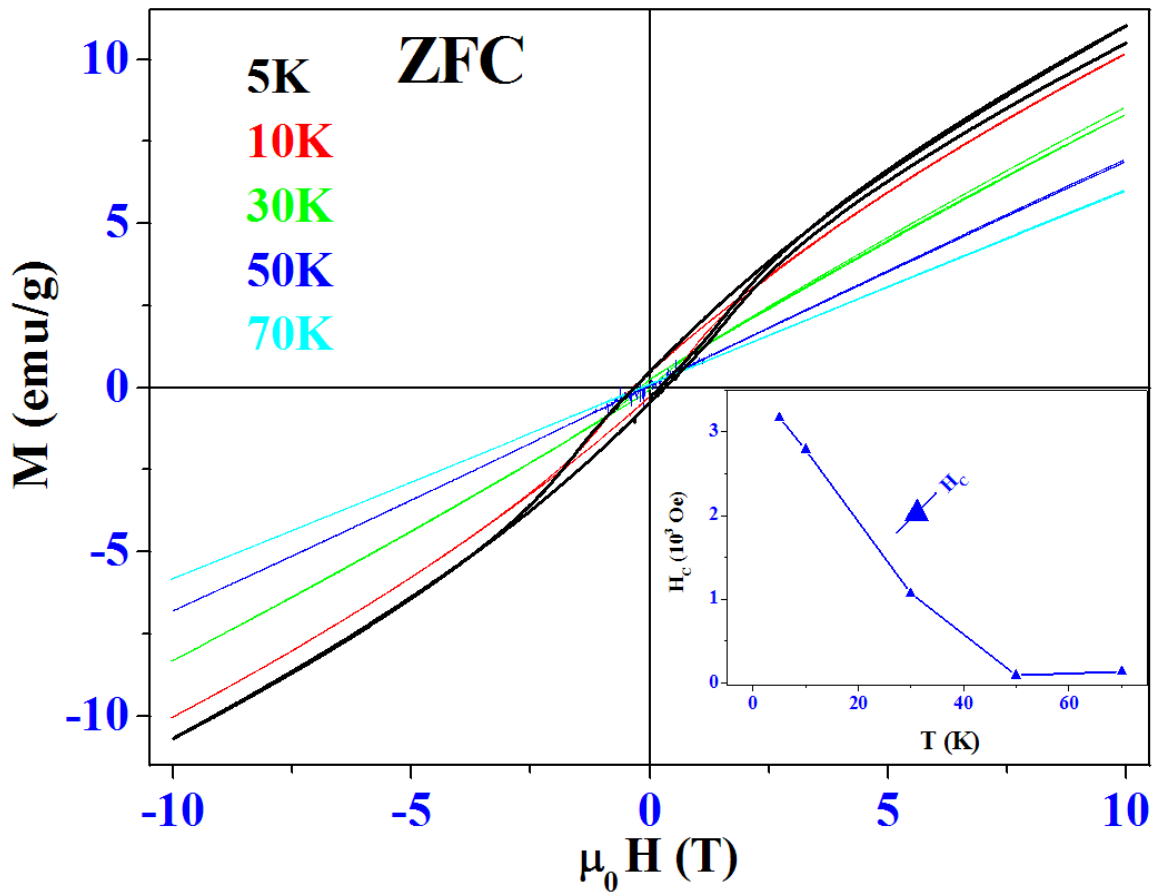

**Fig. S2.** M vs. H curves of α-MnO$_2$ square nanotubes at 5 K, 10K, 30K, 50K, and 70K after ZFC. The inset shows the coercivity field (H$_C$) at different temperature.

Fig. S2 shows M - H hysteresis curves measured at temperatures of 5, 10, 30, 50, and 70 K after ZFC with applied fields up to 90 kOe, while the inset shows the H$_C$ vs. temperature curve. The hysteresis loop recorded after cooling the system in zero external field exhibits a small hysteresis (with a remanent magnetization of 0.48 emu/g), which is symmetrical and centered about the origin. The ZFC M-H loops are point symmetric with a H$_C$ of 3360 Oe at 5 K. The origin of this ferromagnetism at low temperature may be due to the small dimensions of the particles. This is because, for a bulk antiferromagnet, the sublattice magnetizations are fully compensated, resulting in zero net magnetization. Similar behavior for a system of antiferromagnetic nanoparticles has been reported for NiO, [47] α-Fe$_2$O$_3$ [48], and a cubic Co$_3$O$_4$ mesostructure [49]. Different models have been proposed to explain this weak ferromagnetism in small antiferromagnetic nanoparticles. For example, Neel [15] attributed this to the uncompensated spins on the two sublattices. Kodama et al. [47] have proposed a model where the spins in antiferromagnetic nanoparticles yield a multisublattice configuration, indicating that the reduced coordination of surface spins leads to an important change in the magnetic order of the whole particle. The presence of some external magnetic impurities in our sample has been ruled out by means of X-ray photoelectron spectroscopy (XPS) analysis, where no traces of metallic Mn have been detected. No other magnetic impurities have been identified by EDS analysis, either. As for the α–MnO$_2$ nanoribbons with those additional unusual magnetic features mentioned above, which could not be fully interpreted using the existing models, we will supply an interpretation later. First, we need to explore more magnetic features; the exchange bias evaluation may provide more information on the origin of the magnetism.

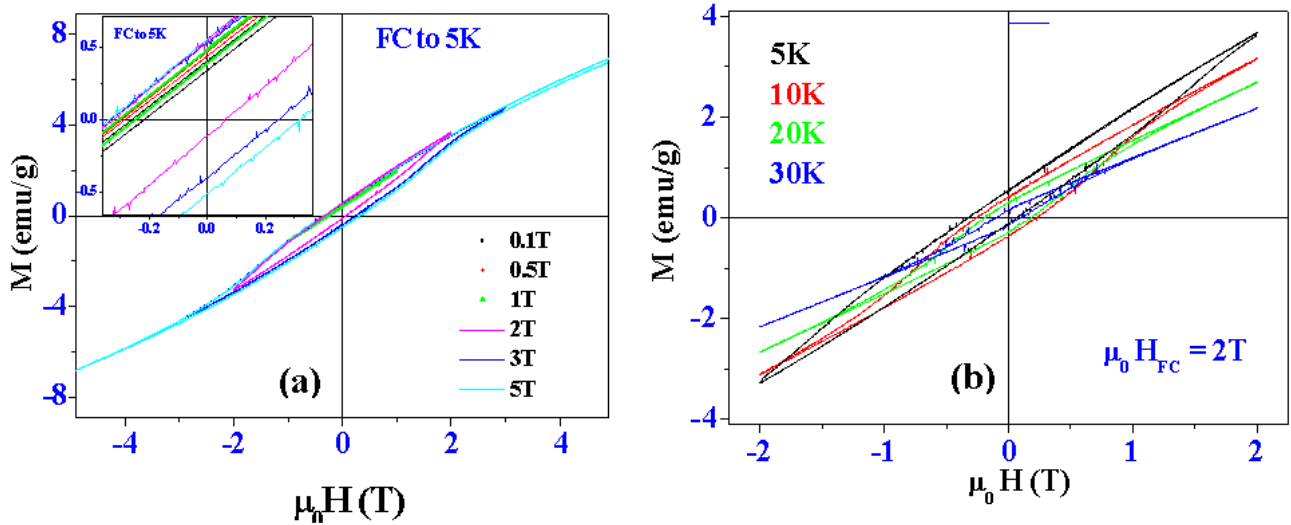

**Fig. S3.** (a) M vs. H curves at 5 K after field cooling under $H_{FC}$ = 0.1, 0.5, 1, 2, 3, 5 T; the inset shows an enlarged region of the M – H curves; (b) M – H curves at 5 K, 10 K, 20 K, and 30 K after field cooling under $H_{FC}$ = 2 T.

Fig. S3(a) shows M - H hysteresis curves measured at 5 K after selected field cooling ($H_{FC}$ = 0.1, 0.5, 1, 2, 3, 4, 5 T), while the inset shows an enlargement of the central part, which displays no significant enhancement of the coercive field, but a significant increase in $H_{HS}$ (magnetization hysteresis loop horizontal shift) at fields lower than 1 T. A shift upward of the hysteresis loop ($M_{shift}$) = $M(H+)$ - $M(H-)$, a training effect accompanied by open loops up to 5 T, and a tunable magnitude of the exchange field via the cooling field are directly observed. Interesting, the sample obviously displays the processing of reverse moments and a critical cooling field (the above mentioned $H^R_C \approx 1$ T); the loop widens and shifts back to the center with a significant increase in the $H_C$, but a decrease in the $H_{HS}$. Moreover, one finds a vertical shift to larger M(H) values. All these notable findings will be discussed below. Usually, a horizontal shift of the hysteresis loop occurs when a ferromagnet (FM) is in contact with an antiferromagnet (AFM) and the entire system is cooled through the Neel temperature of the antiferromagnet [50, 51]. The exchange coupling present at the interface between the FM and the AFM induces a unidirectional anisotropy of the ferromagnetic layer. The strength of the unidirectional anisotropy is measured by the magnetization hysteresis loop horizontal shift field $H_{HS} = (H_{C1} + H_{C2})/2$.

The unusual phenomenon is an indication that there are two exchange couplings that dominate the $H_{EB}$ and that there is an interplay between them, that of the α-MnO$_2$ AFM ($T_N$ about 13 K) substrate exchange coupling with both surface / interface SG-like phases and the step-edge ions when the temperature is lower than about $T_N$, while the SG clusters show a similar exchange coupling with the strongly anisotropic behavior of the FM step-edges when the temperature is higher than $T_N$. This can be explained as follows: by increasing the magnitude of the cooling field (at T < $T_N$), more uncompensated spins (at the interfaces of AFM/SG clusters) are aligned with the magnetic field and rotate. However, this can't explain the appearance of $H^R_C$ behavior, which is due to the strong anisotropy of the step-edges. Rusponi et al. [13] and Kuch et al. [52] suggested that the step-edge atoms can cause 20 times larger magnetic anisotropy due to their large orbital magnetic moment, so we conclude that the dominant contribution to the magnitude of $H_{HS}$ is the step-edge orbital ordering, which needs a 1 T ($H^R_C$) field to rotate it, so when the applied $H_{FC}$ is larger than $H^R_C$, these orderings will be rotated in the applied field direction and cancel parts of the $H_{HS}$, so the $H_{HS}$ start to decreases with a continuing increase in $H_{FC}$, but it can continue to contribute to the $H_C$. Moreover, the magnetic configuration at the AFM / SG-cluster interfaces does not change, so $H_C$ continually increases with the applied $H_{FC}$, but much more slowly. Thus, for the below $H^R_C$ range, with increasing magnitude of the cooling field, more and more frozen-in spins and step-edge orbital orderings are created, and a tunable exchange bias field is obtained. For the above $H^R_C$ range, there is a competition between more frozen-in spins and canceled step-edge orbital ordering (canceled $H_{HS}$), When the frozen-in spins reach saturation, the $H_{HS}$ decreases monotonically with increasing $H_{FC}$, as observed in Fig. S1(a) and Fig. 4(b). Moreover, when the $H_{FC}$ is fixed in the mixed state (two component competition range) at 2T, we observe the unusual $H_{HS}$ and $H_C$ behavior with temperature described above, which can be interpreted as follows: The $H_{HS}$ behavior with increasing temperature can be explained using a simple model, in which both exchange couplings with AFM (in α-MnO$_2$ substrate ribbons) weaken with increasing T to close to $T_N$, and the $H_{HS}$ is reduced sharply. When $T > T_N$, both exchange couplings with AFM are canceled, leaving the exchange coupling of the SG cluster-like phase with step-edge FM, which coupling results in much smaller $H_{EB}$ than the AFM-FM. A characteristic feature of SG-FM coupling is that a positive $H_{HS}$ exists around the conventional blocking temperature ($T_B = T_C$), as shown in the enlarged inset of Fig. 4(b), which was well explained by Rusponi, S et al. [13], in which it was proposed that the additional Zeeman energy provided during the cooling led to the positive exchange bias shift or magnetization hysteresis loop horizontal shift. It seems much more complicated to explain the unusual variation in the $H_C$ behavior with increasing T: the $H_C$ increases at $T < T_N$, and then decreases at $T > T_N$ with T increasing, and the maximum value of $H_C$ appears at the same $T_N$ temperature. These phenomena look similar to those observed by M. Ali et al. [52] in the Co/CuMn system and by T. Gredig et al. [53] in the Co/CoO system, but there are significant differences. Both the Co/CuMn and the Co/CoO systems show an $H_C$ peak and positive $H_{EB}$ around the blocking temperature, but there is unusual behavior of both coercive fields towards the cooling–field direction in the Co/CuMn system, and the interpretation is different.

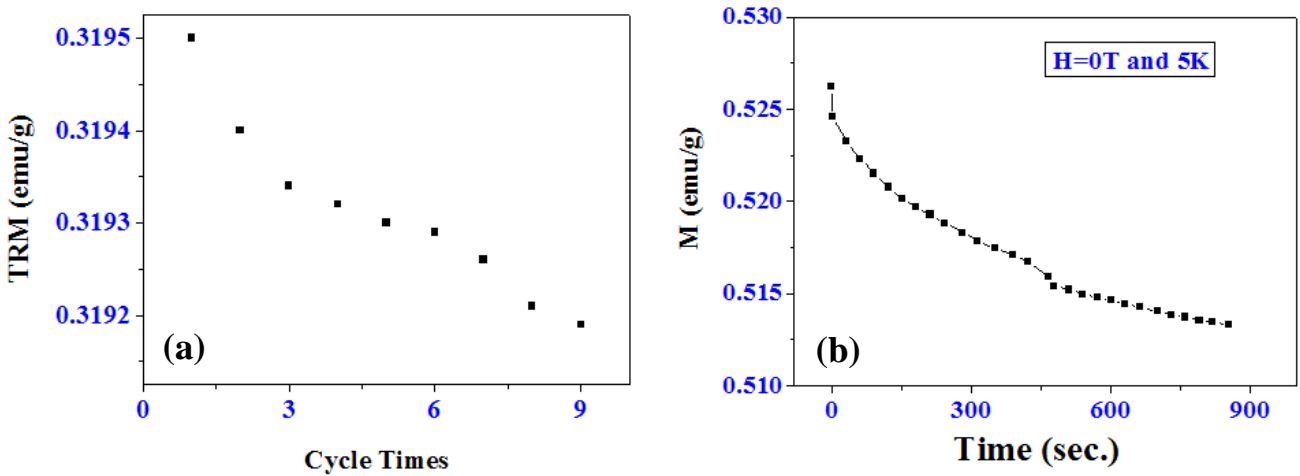

**Fig. S4.** (a) Field cycle dependence of the thermoremanent magnetization (TRM) moment recorded at T = 16 K; (b) TRM measured at 5 K with delay time: the TRM decays slowly and only a decrease of less than 2.6% of the TRM is detected after 900 s.

The magnetism of MnO$_2$ nanostructures seems to only originate from the surfaces or interfaces, as well as the interactions between the nanostructures, so our neutron diffraction measurements (the results not show here) on α–SNTs do not find interesting magnetic behaviours such as are observed by the PPMS measurements, but they thus confirm that the origin of magnetism is from the surfaces and interfaces. The neutron diffraction patterns at different temperatures and fields are shown in the supporting information.

The induced magnetism of the clusters should first present SPM behavior, and SPM really does have a role in the above demonstrated magnetic behavior in α–SNTs. However, the correlation between the SPM behavior of $MnO_2$ that is present in the nanostructures and their sizes is not consistent with the size dependence of magnetic anisotropy in magnetic nanoparticles according to the Stoner-Wohlfarth theory [53 -55]. This is because the microscopic origin of the magnetic anisotropy of the nanostructures stacked from nanoribbons is different from that of the corresponding nanoparticles and is mainly dependent on the step-edges. The magnetic anisotropy is an energy barrier to prevent magnetization from varying from one direction to the other. The blocking temperature is the threshold point of thermal activation to overcome such a magnetic anisotropy and to transfer magnetic nanostructures to the superparamagnetic state. A larger amount of step-edges implies a higher magnetic anisotropy energy, and consequently, a higher thermal energy is required for nanostructures to become superparamagnetic. Therefore, the blocking temperature would increases with an increasing amount of step-edges. The hysteresis in the field-dependent magnetization of the $MnO_2$ nanostructures below the blocking temperature clearly indicates that the magnetic anisotropy serves as an energy barrier to prevent the magnetization orientation of nanocrystals from closely following a switch in the magnetic field direction. The coercivity ($H_C$) represents the required strength of the magnetic field to overcome the magnetic anisotropy barrier and to allow the magnetization of nanocrystals to align along the field direction. The coercivity of a magnetic nanocrystal from the Stoner- Wohlfarth theory.

When the temperature is below the blocking temperature for the given nanocrystals, the required coercivity for switching the magnetization direction of the nanocrystals certainly increases as the magnetic anisotropy increases.

α-$MnO_2$ is a 2 × 2 tunnel structure antiferromagnetic material, so it is possible that the insertion of large cations into the tunnel could lead to changes in its magnetic properties. On the one hand, the reduction in the oxidation state of Mn (mixed valence of $Mn^{3+}$ and $Mn^{4+}$) is necessary in order to compensate the charge of the introduced large cation, [2, 3] which could influence the magnetic coupling between Mn cations. On the other hand, the distribution of the $Mn^{3+}$ and $Mn^{4+}$ cations should closely relate to the distribution of the intercalated cations, which in turn may cause a change in the magnetic ground state. When the particles are down to nanosize, as in our $MnO_2$ nanostructures case, in addition to the surface / interface oxygen vacancies, which generate additional $Mn^{3+}$ ions at the step-edges, there is a more complicated distribution of the $Mn^{3+}$ and $Mn^{4+}$ cations in the surface / surface interfaces and the step edges, so all the $MnO_2$ nanostructures present weak ferromagnetism. However, to turn to the unusual magnetic phenomena in α-$MnO_2$ square nanotubes, a large cation (here $K^+$) introduced into the tunnel cavity causes a mixture of $Mn^{3+}$ and $Mn^{4+}$ in the sample based on the valence balance, and this is the natural origin of the ordinary magnetism, but it cannot explain the natural origin of the unusual magnetic phenomena, since other $MnO_2$ nanostructures, especially δ−NF, have the same amount of $K^+$ ions and show the same valence states as α−SNT (see Fig. 1(b) and Fig. S2 containing the XPS analysis), so the origin of the unusual magnetic phenomena in α−SNT seems mainly to lie in the unique square morphology.

By careful XRD refinement calculations, we have detected a small expansion of the unit cell parameters $a = 0.985$ nm and $c = 0.286$ nm for our α−SNT sample, which can be ascribed to the packing of $K^+$ ions into the tunnel cavity of α−$MnO_2$, as illustrated in Fig. 1(b), and nanoscale effects. These results indicate that we can rule out the $K^+$ ions as the main reason for the unusual magnetic phenomena, but they really do enhance the magnetism and the unusual phenomena. It is well known that magnetic $Mn^{3+}$/$Mn^{4+}$ ions are triangularly arranged, and there is a triangular arrangement of magnetic moments with the Mn−Mn distance about 0.294 nm in our α−SNT sample. We can concentrate on the unique stacked nanoribbon square morphology with a large amount of step-edges and corners, which should be the main reason.

Detailed magnetic measurements and analysis are presented, on α-MnO$_2$ square nanotubes (α–SNT). Magnetization measurements show that α–SNT exhibits superferromagnetism at temperatures higher than the antiferromagnetic transition temperature. The unusual magnetic properties of nanotubes include the interactions (in the form of coupling and competition) involving ground state or magnetic behaviors from the nanotube ribbons, the surface clusters, and the step-edges. Conventional magnetic measurements hide the AFM transition; however, magnetization hysteresis loop horizontal shift studies show unusual and significant $H_{HS}$ phenomena. The irreversible magnetization contributions are due to two exchange couplings: ribbon surface clusters, which behave as a superferromagnetic system, and ribbon step-edges, which behave as very strong anisotropic features and have ferromagnetism with high anisotropy energy, which possible the oringal of large unquenched orbital moment.

The magnetic interaction and coupling results show that AFM ribbons undergo a strong exchange interaction with SPM clusters and FM step-edges. They present significant $H_{HS}$ at temperatures below $T_N$ under lower than critical cooling fields ($H^R_C$). AFM ribbons, SFM clusters, and FM step-edges have a weak exchange interaction with each other, present low $H_{HS}$ at temperatures above $T_N$ and at fields higher than the critical cooling field ($H^R_C$), and even invert the $H_{HS}$ at temperatures close to and higher than $T_C$. The $H_{HS}$ is significant for electronic devices and is a more effective tool to study the nature of nanomagnetism, the microscopic origin of anisotropy energy, and the hinted magnetic phase transition.

Microscopic origin studies on the magnetism of α-MnO$_2$ nanotubes indicate that the magnetic moments consist of three parts: (i) The coupling between $Mn^{4+}$- $Mn^{4+}$ pairs (antiferromagnetic) in the α-MnO$_2$ based ribbons is the origin of spin ordered AFM; (ii) Oxygen vacancies and uncompensated electrons of step-edge $Mn^{3+}$ ions cause $Mn^{3+}$ - $Mn^{4+}$ geometrical frustrations (GFs), which mainly result in a ferromagnetic orbital moment on $Mn^{3+}$. These give rise to well aligned long range ordered ferromagnetism and high anisotropy energy, which may also create skyrmion lattice-like vortex magnetic moments in the nanoribbon step-edges, in addition to well-aligned orbital ordered magnetization antiparallel to the applied field, leading to our observations of unusual magnetic phenomena, the negative $\Delta M$ (= $M_{FC} - M_{ZFC}$) values, the feature of $H^R_C$, and enormous uniaxial magnetocrystallographic anisotropy. These factors suggests the experimental observation of the presence of an unusual high unquenched orbital moment, which is responsible for the antiparallel magnetization and results in the abnormal $H_{HS}$ phenomena, e.g., $H_{HS}$ appears as a peak in $H_{FC}$ at $H^R_C$. (iii) There is strong inter-cluster interaction of the surface heterostructural clusters and K ions in tunnel MnO$_2$ clusters, and those clusters participate in short distance ordering and long distance interaction with each other as well, and hence are a source of superparamagnetic, spin-glass-like behaviors and even a charge source of orbital ordering. Furthermore, x-ray magnetic circular dichroism (XMCD), x-ray absorption spectroscopy (XAS), or inelastic neutron scattering measurements will be conducted in the future to separate the spin and orbital moment contributions and identify the coupling and competition of the electron spin and orbital moments, and the different ground states.

**References.**